\documentclass[sn-aps, Numbered, iicol]{sn-jnl}

\usepackage{graphicx}%
\usepackage{multirow}%
\usepackage{amsmath,amssymb,amsfonts}%
\usepackage{amsthm}%
\usepackage{mathrsfs}%
\usepackage[title]{appendix}%
\usepackage{xcolor}%
\usepackage{textcomp}%
\usepackage{manyfoot}%
\usepackage{booktabs}%
\usepackage{algorithm}%
\usepackage{algorithmicx}%
\usepackage{algpseudocode}%
\usepackage{listings}%
\usepackage{braket}

\usepackage{MnSymbol}
\usepackage{array}
\newcolumntype{C}[1]{>{\centering\arraybackslash}p{#1}}
\usepackage{makecell}
\usepackage{boldline} 
\usepackage{tabularx}
\usepackage{subcaption}

\begin{document}

\title[Quantum kernel and HHL-based support vector machines for multi-class classification]{Quantum kernel and HHL-based support vector machines for multi-class classification}

\author[1]{\fnm{Gabriela} \sur{Pinheiro}}\email{gpinheiro2812@gmail.com}

\author*[2]{\fnm{Donovan, M.} \sur{Slabbert}}\email{donovanslab@mweb.co.za}

\author[1]{\fnm{Luis} \sur{Kowada}}\email{luis@ic.uff.br}

\author[3,4]{\fnm{Francesco} \sur{Petruccione}}\email{petruccione@sun.ac.za}

\affil[1]{\orgdiv{Instituto de Computação}, \orgname{Universidade Federal Fluminense}, 
\orgaddress{\city{Niterói}, \state{Rio de Janeiro}, \country{Brazil}}}

\affil*[2]{\orgdiv{Department of Physics}, \orgname{Stellenbosch University}, \city{Stellenbosch}, \postcode{7600}, \country{South Africa}}

\affil[3]{\orgdiv{School of Data Science and Computational Thinking}, \orgname{Stellenbosch University}, \city{Stellenbosch}, \country{South Africa}}

\affil[4]{\orgname{National Institute of Theoretical and Computational Sciences (NITheCS)}, \city{Stellenbosch}, \country{South Africa}}

\abstract{We compare two quantum approaches that use support vector machines for multi-class classification on a reduced Sloan Digital Sky Survey (\texttt{SDSS}) dataset: the quantum kernel-based QSVM and the Harrow-Hassidim-Lloyd least-squares SVM (HHL LS-SVM). Both one-vs-rest and two-step hierarchical classification schemes were implemented. The QSVM involves angle encoding of ten features, two unitary operator blocks consisting of rotational operator gates, and a projective measurement that projects the final state to the zero state. The HHL-based method involves solving a system of linear equations using the HHL algorithm and using the solution in a support vector machine approach. The results indicate that the QSVM outperforms HHL LS-SVM in most cases. HHL LS-SVM performs somewhat competitively in selected cases, particularly when isolating galaxies (majority), however, it also performs poorly in others, especially when isolating QSOs (minority). Comparisons with classical SVMs confirm that quantum and classical methods achieve broadly similar performance, with classical models performing slightly ahead overall. Scaling analysis reveals a trade-off: QSVM performance suffers from quadratic scaling with the number of samples and features, but benefits from explicit feature representation during training, while HHL LS-SVM scales essentially constantly, with moderate fluctuations, but suffers from limited representative elements. The HHL-based method is also highly noise-sensitive. These results suggest that QSVM performs better overall and will perform better on current hardware as well, but that the more efficient scaling of HHL LS-SVM makes it a useful option for larger datasets with many samples, especially if we move past the NISQ era.}


\maketitle

\clearpage 

\section{Introduction}\label{Introduction}

The interdisciplinary field of quantum machine learning (QML) is at the intersection of quantum computing and classical machine learning \cite{schuld2021machine}. Quantum technologies are being improved, and as quantum computers continue to advance, there is growing interest in leveraging the properties of quantum systems (such as superposition, entanglement, and quantum parallelism) to enhance machine learning algorithms beyond classical capabilities. The goal of QML is to target both the acceleration of computationally intensive tasks and to uncover possible novel approaches to learning from data through quantum algorithms \cite{biamonte2017quantum, schuld2015introduction}. Developments in quantum hardware, along with theoretical progress in quantum algorithms, have laid the foundation for exploring whether it is practically viable and whether there are any advantages to quantum-enhanced learning models \cite{preskill2018quantum, arute2019quantum}. There are significant challenges to overcome before this can become a reality. The current noisy intermediate-scale quantum (NISQ) \cite{preskill2018quantum, kandala2017hardware} supplies only fragile quantum states due to noise and decoherence. The difficulty of efficiently encoding large-scale classical data into quantum circuits is also a major obstacle \cite{havlivcek2019supervised}. QML can play an important role in the future of data-driven science, with possible applications covering multiple fields and disciplines, but only if these challenges are addressed or mitigated through advances in quantum error correction \cite{Lidar_Brun_2013, nielsen2010quantum}, encoding techniques \cite{havlivcek2019supervised, schuld2019quantum}, and hybrid algorithm design \cite{cerezo2021variational, mitarai2018quantum}.

Multi-class classification is a particularly challenging and interesting application for QML. This is especially true in scientific domains such as astronomy, which is data-intensive. Large-scale surveys such as the Sloan Digital Sky Survey (\texttt{SDSS}) \cite{york2000sloan} generate large amounts of high-dimensional data that comprise millions of mostly unlabeled samples. These surveys contain data of multiple astrophysical objects such as stars, galaxies, and QSOs (quasars). Although QML may offer potential advantages in learning complex decision boundaries \cite{rebentrost2014quantum, schuld2019quantum}, its application to such datasets is hindered by the infeasibility of encoding large volumes of classical data into quantum circuits \cite{havlivcek2019supervised, schuld2021supervised}. The high dimensionality of the feature space, coupled with the need to process many samples and build representative model architectures for complicated data, places severe constraints on circuit width and depth, particularly in the NISQ era, where available qubits are limited and noise rapidly degrades circuit fidelity \cite{preskill2018quantum, kashyap2025advances}. This makes efficient data encoding \cite{rath2024quantum} and scalable or clever quantum architectures \cite{qin2023review} essential prerequisites for applying QML to real-world multi-class classification problems in astrophysics.

Our primary goal was to explore the application of two specific QML techniques to multi-class classification in high-dimensional datasets, such as those found in astrophysics, and to compare them directly. We aimed to find out which of these methods best circumvented the challenge posed by the number of samples large-scale datasets, such as the \texttt{SDSS}, typically has. Processing and classifying such data efficiently require quantum algorithms that can scale with data size while also managing circuit complexity and noise limitations inherent to current quantum hardware. Multi-class classification in the context of QML has been explored before \cite{bokhan2022multiclass, PhysRevA.111.062410, pillay2024multi, cappelletti2020polyadic, yun2022projectionvaluedmeasurebasedquantum, zhang2022quantumalgorithmneuralnetwork}. 

We focus on two quantum classification approaches: the quantum-enhanced support vector machine (QSVM) \cite{havlivcek2019supervised, liu2021rigorous, huang2021power}, which leverages quantum kernel evaluation, and a HHL-based least-squares support vector machine (HHL LS-SVM) \cite{pinheiro2025}, which uses a quantum linear systems algorithm for training. By analyzing and comparing these methods, we seek to advance the understanding of the practical deployment of these quantum classifiers or models and to determine whether they are capable of distinguishing among multiple astrophysical classes with hopefully improved efficiency and accuracy.

The first method we used in the comparison, the QSVM, is a foundational algorithm in QML, and it is designed to exploit quantum kernels for classification. The QSVM uses a quantum kernel, which is also a similarity measure between samples based on their quantum state representations. This allows non-linear decision boundaries in high-dimensional feature spaces to be exploited \cite{huang2021power}. Once the kernel matrix is computed, a classical SVM is employed to perform linear separation in the induced feature space. This approach is usually used for binary classification, but can be extended to handle multi-class classification through strategies such as one-vs-rest and one-vs-one schemes \cite{weston1999support, hsu2002comparison}. These approaches involve training multiple binary QSVM classifiers and cleverly aggregating their outputs to predict multi-class labels. In large-scale datasets with more than two classes and complex feature distributions, such as those found in astrophysics or surveys like \texttt{SDSS}, this extension becomes computationally costly due to the need for repeated kernel evaluations \cite{hsu2002comparison}, both for the number of samples and for the number of unique combinations between two classes for each binary classifier. QSVM remains a straightforward, but effective approach for structured classification problems, but only if quantum hardware and encoding strategies can scale to support its demands. Kernel methods, with QSVMs being prominent examples, have been applied many times in different contexts \cite{blank2020quantum, innan2023enhancing, incudini2024automatic, akpinar2024evaluating}, where previous work focused on their direct application to astronomical datasets \cite{slabbert2024pulsar, slabbert2024hybrid}.

The second method, the HHL LS-SVM, leverages the quantum linear systems algorithm developed by Harrow, Hassidim, and Lloyd \cite{harrow2009quantum} to reformulate the classical SVM training task as a quantum linear system problem. Representing the kernel-based optimization objective of SVMs as a linear system, it allows the HHL algorithm to be employed to solve for the support vector coefficients in quantum superposition. This offers a potential speedup under ideal conditions \cite{rebentrost2014quantum}. Originally applied to the HTRU-2 pulsar classification dataset \cite{pinheiro2025}, the HHL LS-SVM approach presents a viable option for extending quantum-enhanced classification to broader domains that include multi-class classification problems. Similarly to the QSVM, multi-class classification with HHL LS-SVM typically involves decomposing the task into multiple binary classifiers using schemes such as one-vs-rest (OvR) or one-vs-one (OvO), which increases computational demands by increasing the number of SVMs used, but also enables the method to address more than two classes.

We implement both methods by using a one-vs-rest multi-class scheme, where three sets of binary classifiers are used and predictions are made by aggregating the probability outputs from each classifier. We also implemented a two-step hierarchical classification approach to address the multi-class classification problem. The method decomposes the three-class problem into two sequential binary classification tasks using a one-vs-rest strategy: the first classifier separates one class from the rest, and the second classifier then distinguishes between the remaining classes. This two-step process reduces computational cost, as compared to traditional one-vs-rest or one-vs-one methods, by only requiring two binary SVMs. This enables multi-class classification through a set of binary classifiers without needing one for each pair of classes. It is in essence a more targeted use of quantum resources and the hope is that any performance decreases, as compared to traditional approaches, will be limited. We explore multiple variations of this approach by alternating which class is selected in the first step, exploring its flexibility and applicability to a real-world astrophysical dataset with three classes.

The results given are only for simulated runs. Real device runs were considered, specifically for the HHL LS-SVM, but since the HHL algorithm is really susceptible to noise \cite{harrow2009quantum, marfany2024identifying, phillips2024detailed, zaman2022study}, the results were significantly worse. This was done to isolate algorithmic performance from hardware noise and limitations. The question is posed: How well can these methods be expected to perform if noise were absent? The objective is then to benchmark and directly compare two quantum approaches to support vector machines. While QSVMs have been applied to various binary and extended multi-class problems, the HHL LS-SVM has not yet been applied to multi-class classification. This study aims to provide a look at how such an approach performs in a higher-dimensional, more data-intensive setting, which also offers insight into the practical viability and scaling behaviour of the HHL LS-SVM in contrast to the more established QML technique.

This paper is structured as follows. Section \ref{Theory} is the theoretical background, that includes detailed discussions of the QSVM and the HHL-based method. Section \ref{Methodology} describes the methodology, covers the Sloan Digital Sky Survey dataset, explains any preprocessing steps, and has detailed descriptions of the quantum and classical classification methods used. This includes the one-vs-rest and two-step SVM approaches. Section \ref{Results} shows and discusses the experimental results and performance comparisons across the different classification approaches. Finally, Section \ref{Conclusion} concludes the paper with a brief conclusion and outlook.

\section{Theory}\label{Theory}

\subsection{Quantum-Enhanced Support Vector Machines}
\label{sec:QSVM}

Support vector machines (SVMs) are supervised learning models designed to find the maximal-margin hyperplane that separates data into two classes \cite{cortes1995support}. If the data is not linearly separable in the input space, a feature map $\phi(\vec{x})$ is used to map the data to a higher-dimensional space, called the feature space, where linear separation becomes possible. SVMs, similar to other kernel methods, benefit from the kernel trick \cite{boser1992training, scholkopf1999advances}. Instead of explicitly computing the full feature map every time, it is possible to evaluate inner products or kernel values in feature space simply by using a kernel function:

\begin{equation}
    K(\vec{x}, \vec{x}') = \langle \phi(\vec{x}), \phi(\vec{x}') \rangle.
\end{equation}

Quantum-enhanced support vector machines (QSVMs) extend this framework by first encoding two classical input feature vectors from two separate samples into quantum states via a quantum feature map \cite{havlivcek2019supervised, schuld2019quantum}. This is implemented as a quantum circuit. The input vectors $\vec{x}$ and $\vec{x}'$ are mapped to quantum states $\ket{\phi(\vec{x})}$ and $\ket{\phi(\vec{x}')}$ in Hilbert space, and the kernel function is defined by the squared inner product (fidelity) between them:

\begin{equation}
    K(\vec{x}, \vec{x}') = |\langle \phi(\vec{x}) | \phi(\vec{x}') \rangle|^2.
\end{equation}

In other words, this quantum kernel is the similarity between two inputs after they have been mapped into quantum state space and is estimated using quantum measurements on encoded states that represent sample feature vectors. The complete kernel matrix is constructed by computing this quantity for all pairs of training samples. A classical SVM is trained or fitted using this matrix.

A common choice of feature map or state encoding for kernel methods is angle encoding, where each component of the input or feature vector $\vec{x} = (x_1, x_2, \dots, x_n)$ is encoded using a rotation on a single qubit with the feature as the rotational parameter. This is shown by:

\begin{equation}
    U(\vec{x}) = \bigotimes_{j=1}^{n} R_Y(x_j),
\end{equation}

\noindent where $R_Y(x_j) = \exp(-i x_j \sigma_Y / 2)$ is a rotation around the $y$-axis on the Bloch sphere for the $j$-th qubit. The resulting state is a unitary operator representing all the y-rotations on each qubit as:

\begin{equation}
    \ket{\phi(\vec{x})} = U(\vec{x}) \ket{0}^{\otimes n}.
\end{equation}

To compute the kernel value $K(\vec{x}, \vec{x}')$ the inverse encoding, or complex conjugate transpose of the first angle encoding, is applied using the feature vector of a second sample. This is denoted as $U^\dagger(\vec{x}')$ and is applied to the state $U(\vec{x})\ket{0}^{\otimes n}$. Measurement involves a projector matrix, where the $z$-expectation value of the state projected to the density matrix of the all-zero state is measured. This value is also the squared overlap, or kernel value:

\begin{equation}
    K(\vec{x}, \vec{x}') = |\langle 0^{\otimes n} | U^\dagger(\vec{x}') U(\vec{x}) | 0^{\otimes n} \rangle|^2,
\end{equation}

\noindent where the projector matrix is simply:

\begin{equation}
    \rho = \ket{0}^{\otimes n} \bra{0}^{\otimes n}.
\end{equation}

In other words, this is estimated by preparing the state $U^\dagger(\vec{x}') U(\vec{x}) \ket{0}^{\otimes n}$ and computing the probability of measuring the zero state.

\begin{figure}[h]
    \centering
    \includegraphics[width=0.7\columnwidth]{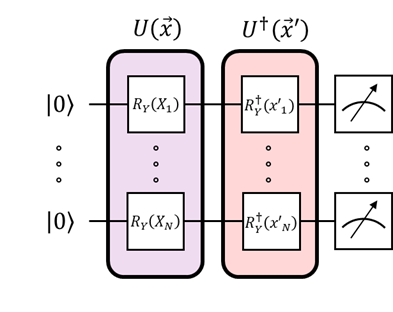}
    \caption{The quantum circuit of the quantum kernel used for the implementation of the QSVM using a projector measurement. Each feature in a feature vector is encoded using y-rotation gates in the first unitary. The second unitary is the complex conjugate transpose of the first, but applied to a second feature vector. The number of qubits equal the number of features, with two angle encoding unitaries before measurement. Measurement involves a projector measurement. The z-expectation value of the state is measured, projected to the zero-state density matrix.}
    \label{fig:fig1}
\end{figure}

This measurement-based approach avoids the need to use ancilla qubits in SWAP tests \cite{MariaSchuld2021}. This reduces circuit complexity, which is a critical consideration when using NISQ devices; however, the evaluation of the kernel matrix still scales quadratically with the number of samples, requiring $O(N^2)$ circuit executions for a dataset of size $N$, which can be costly. The quantum circuit is illustration in \ref{fig:fig1}.

QSVMs offer some advantages. They avoid barren plateaus \cite{mcclean2018barren} that often affect variational quantum circuits \cite{schuld2021supervised}, they are conceptually simple, and seamlessly integrate with classical SVM theory. We therefore employ QSVMs to classify astronomical objects in a higher-dimensional dataset and we evaluate their use in two-step and one-vs-rest multi-class classification schemes.

\subsection{HHL LS-SVM}\label{HHL LS-SVM}

\begin{figure*}[t]
    \centering
    \includegraphics[width=\textwidth]{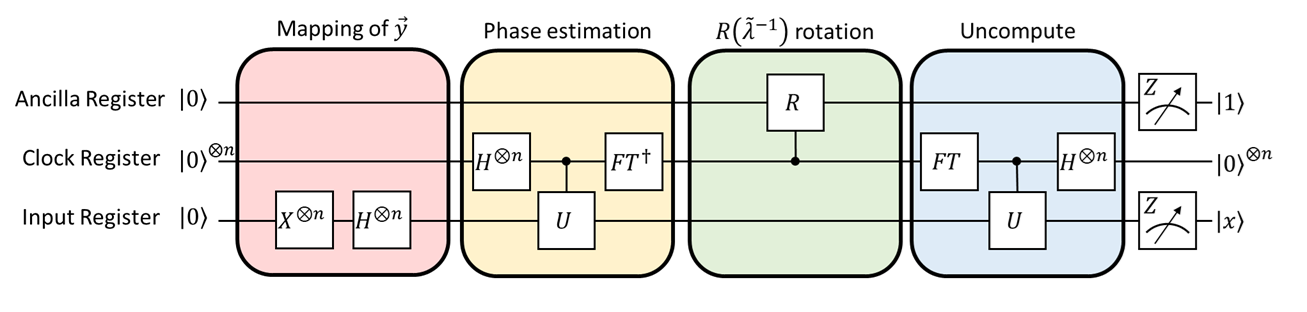}
    \caption{The quantum circuit for the implementation of the quantum HHL LS-SVM. The circuit is divided in four parts: it begins by mapping the $\vec{y}$ vector into the input register using a superposition of the state $\ket{-}$. This is followed by quantum phase estimation used to store the kernel matrix in the clock register. A series of controlled $R_Y$ rotations is performed at the ancilla qubit and the circuit is finalized with an inverse quantum phase estimation, removing the matrix from the clock register and is known as uncomputation. The measurements will be performed on the two registers with remaining information. With only the results where the ancilla qubit is measured as "1" being considered, the solution will be stored in the amplitudes of the input register. The ancilla register uses 1 qubit, the input register uses $\lfloor\log_2n\rfloor +1$ qubits, and the clock register uses $\lfloor\log_2\kappa\rfloor + 1$ qubits. This results in a total of $\lfloor\log_2n\rfloor + \lfloor\log_2 \kappa\rfloor + 3$ qubits.}
    \label{fig:fig2}
\end{figure*}

The separating hyperplane for a support vector machine is defined by $\Vec{w} \cdot \Vec{x} - b = 0 $. The goal is to maximize the margin $\frac{2}{||\Vec{w}||}$, which is equivalent to minimizing $\frac{\left \| \Vec{w} \right \|^2}{2}$ under the constraint:

\begin{equation}
    y_j(\Vec{w} \cdot \Vec{x_j}+b)\geq 1,
\end{equation}

\noindent where $\Vec{x_j}$ is an element of the dataset and $y_j$ is the class label.

Taking into account the training error, represented by $e_j$, and adding it to the equation, the inequality restriction can be transformed into the equality:

\begin{equation}
    y_j(\Vec{w} \cdot \Vec{x_j}+b) = 1 - e_j.
\end{equation}

This allows for a least-squares reformulation of the original problem \cite{suykens1999least}. This reformulation, known as LS-SVM, transforms the SVM problem into a resolution of a linear system.

With the transformation, the use of quantum linear system algorithms was proposed to solve the LS-SVM system presented in Equation \eqref{eq:f_original} \cite{rebentrost2014quantum}:

\begin{equation}
\label{eq:f_original}
\begin{split}
    F
    \begin{pmatrix}
    b' \\ 
    \Vec{a}
    \end{pmatrix}
    \equiv
    \begin{pmatrix}
    0 & \Vec{1}^T \\
    \Vec{1} & K + \gamma^{-1}I
    \end{pmatrix}
    \begin{pmatrix}
    b' \\ 
    \Vec{a}
    \end{pmatrix}
    &=
    \begin{pmatrix}
    0 \\ 
    \Vec{y}
    \end{pmatrix},
\end{split}
\end{equation}

\noindent where $K$ is the kernel matrix, a $n \times n$ matrix. The dataset information will be indirectly inserted by $n$ representative elements using the kernel function $K_{ij} = k(\Vec{x_i},\Vec{x_j}) = \Vec{x_i}\cdot\Vec{x_j}$, where $n$ is the dimension of the linear system. In this work, the representative elements of the kernel matrix are classically calculated using the average values of the selected features for each class. $\Vec{y} = (y_0, y_1,...y_{n-1})$, $\Vec{1}^T = (1,...,1)$, $I$ represents an identity matrix, and $\gamma$ is a user-defined variable that determines the relative weight between the training error and the objective. This results in a $(n+1)\times(n+1)$ matrix $F$ and transforms the SVM generation into a linear system in terms of $b'$ and $\Vec{a}$, where the class of a new element can be defined by Equation \eqref{eq:y_original}.

\begin{equation}
\label{eq:y_original}
y_i=
\left\{\begin{matrix}
+1\ \textrm{if}\ \sum\limits_{j=0}^{n-1} a_ik(\Vec{x_i},\Vec{x_j}) + b' \geq 0,
\\ 
-1\ \textrm{if}\ \sum\limits^{n-1}_{j=0}{a_ik(\Vec{x_i},\Vec{x_j})} + b' < 0.
\end{matrix}\right.
\end{equation}

A possible quantum algorithm to be used is the Harrow-Hassidim-Lloyd (HHL) \cite{harrow2009quantum} algorithm. The HHL algorithm is a quantum algorithm designed to solve systems of linear equations, such as $A\Vec{x} = \Vec{b}$, where $\Vec{b}$ is a vector and $A$ a matrix. For matrices with a sufficiently small conditional number $\kappa$, calculated using the ratio between the largest and smallest eigenvalues, it is claimed that the HHL has an exponential gain in complexity compared to the classical linear system algorithms. The resulting HHL quantum state is proportional to Equation \eqref{eq:quantum_linear_system_inverse}, and the solution is stored in the amplitudes of a quantum register.

\begin{equation} 
 \label{eq:quantum_linear_system_inverse}
        A^{-1}|b\rangle = |x\rangle
\end{equation}

Since it would only be possible to obtain a normalized version of the solution, a reduction of the linear system in Equation \eqref{eq:f_original} was proposed \cite{yang2019support}. The reduction takes place by fixing $b' = 0$ and removing the line and column that would originally be multiplied by it, resulting in Equation \eqref{eq:reform}. After removing the scalar value, the normalized version of the remaining variable $\Vec{a}$ has enough information to generate the desired LS-SVM. 

 \begin{equation}\label{eq:reform}
    F
    \begin{pmatrix}
    \Vec{a}
    \end{pmatrix}
    \equiv
    \begin{pmatrix}
     K + \gamma^{-1}I
    \end{pmatrix}
    \Vec{a}
    =
    \Vec{y}
    \end{equation}

Since the data information is stored in the kernel matrix and not directly in the circuit as with the QSVMs, the use of representative elements generates a reduction in the linear system size. This allows the HHL algorithm to be implemented using current NISQ era devices and in simulations. Another consequence of the use of representative elements is that the amount of data in the dataset does not interfere with the circuit size, allowing it to process larger datasets faster than other QML methods. The quantum circuit for the implementation of the HHL LS-SVM is illustrated in \ref{fig:fig3}.

\section{Methodology}\label{Methodology}

\subsection{Sloan Digital Sky Survey Dataset}\label{SDSS}

We apply both methods to astronomical survey data from a reduced version of the Sloan Digital Sky Survey (\texttt{SDSS}) \cite{clarke2020identifying, a_o_clarke_2020_3768398}. We focus on a 10-dimensional reduced feature set selected for their relevance and discriminative power as determined by the creators of the dataset \cite{clarke2020identifying}. These features are: \texttt{psf\_u}, \texttt{psf\_g}, \texttt{psf\_r}, \texttt{psf\_i}, \texttt{psf\_z}, \texttt{w1}, \texttt{w2}, \texttt{w3}, \texttt{w4}, and \texttt{resolvedr}. The labels were assigned using previous machine learning techniques. The labels are as follows:

\begin{equation*}
\text{STAR} \rightarrow 0, \quad \text{GALAXY} \rightarrow 1, \quad \text{QSO} \rightarrow 2.
\end{equation*}

Note that QSO refers to quasi-stellar objects. The dataset was cleaned by discarding any unclassified data, with $1,549,729$ elements remaining.

We evaluated the performance of the methods using commonly used metrics: accuracy, precision, recall, and F1 score \cite{bishop2006pattern}. In each case, these values need to be as close as possible to one to indicate better performance. These values are calculated for both the one-vs-rest approach and the two-step hierarchical approach already mentioned. Finally, we include a plot that illustrates how the number of required operations scales as the number of samples increases for each method.

\subsection{QSVM details}

The quantum kernel employed is a simple two-gate quantum kernel, where we apply angle encoding with y-rotation gates to each qubit in the first layer and the adjoint or complex conjugate transpose of y-rotations to the second layer. The quantum device used for simulations is the \texttt{pennylane} \texttt{lightning.qubit} simulator \cite{bergholm2018pennylane}. Note that the initial state is initialized to the zero state. The number of qubits used is equal to the number of features in the reduced \texttt{SDSS} dataset. The parameter by which we rotate the qubit is equal to the feature in the feature vector corresponding to that qubit, after scaling to $[0, \pi]$. Kernel values are calculated for a pair of samples. One set of features are encoded by the first unitary and the second set of features are encoded using the second adjoint unitary. Kernel value calculation involves a projective measurement, that projects the state down to the zero matrix projector. This matrix has all zero entries, except for the top left entry, which is a one. This can be thought of as the density matrix representing the zero state.

The kernel values are calculated pair-wise for each pair of samples, and are then fed into a standard classical SVM. This is done through \texttt{sckit-learn}. This hybrid quantum-classical workflow allows us to take advantage of quantum feature spaces while utilizing established classical SVM fitting for training. We fit and test the SVM to precisely 300 samples each. This is done to mitigate the quadratic scaling common to SVMs and many other kernel methods as much as possible. We also introduce class weights, calculated as the inverse of each class's relative frequency, normalized by the number of classes. The weight for class $i$ is given by

\begin{equation}
w_i = \frac{N}{C \times n_i},
\end{equation}

\noindent where $N$ is the total number of samples, $C$ is the total number of classes, and $n_i$ is the number of samples in class $i$. This weighting scheme assigns higher weights to less frequent classes to balance their influence during training. This implies that the QSVM has the benefit of knowing beforehand how many samples are in each class. This is not always the case in real-world applications and should be noted. We repeated the runs for randomly sampled sets of 300 training and 300 testing samples a number of times and took the average and standard deviation to be the uncertainty. 

\subsection{HHL LS-SVM details}\label{HHL LS-SVM details}

With the training set being indirectly represented by the kernel matrix, not only does this leave all the preprocessing of the data to be done classically, but it also allows for the use of the full dataset. This is because the number of executions and the circuit size are not correlated with the number of samples used. This can also be affirmed for the testing phase, performed classically using the results measured by the circuit. This favorable scenario removes the requirement of undersampling, since the complete dataset is used during the HHL experiments.

An interesting consequence of creating a kernel matrix using representative elements for unbalanced scenarios, which is the case with the \texttt{SDSS}, is that both classes are equally represented in the system and could compensate for the under-representation of a class in the dataset.

A characteristic of the HHL algorithm is the variance in the number of required qubits for each linear system represented, which tends to generate circuits that are larger than the QSVM circuits. Since the number of qubits that can be simulated is still limited, the number of representative elements was fixed to two, which limits the amount of information fed into the circuit. This tends to result in lower performance compared to the QSVMs.

\subsection{One-vs-Rest SVM}

One way to handle multi-class classification tasks with support vector machines (SVMs) is the one-vs-rest (OvR) approach. Separate binary classifiers are trained for each class against all other classes combined. For a three-class problem with stars, galaxies, and QSOs, this results in a total of three binary SVMs: one trained to distinguish between stars vs. non-stars, another for galaxies vs. non-galaxies, and a third for QSOs vs. non-QSOs.

During testing, a test sample is passed through all three classifiers. Each classifier produces a binary decision, effectively indicating whether the sample belongs to the class it was trained to recognize or to the ``rest". The final class label is determined by selecting the classifier with the highest confidence or decision score. For example, let the three OvR classifiers output scores of 0.8 for galaxy vs. rest, 0.6 for star vs. rest, and 0.4 for QSO vs. rest on a given test sample. The predicted class would be galaxy, because it has the highest score. In cases where more than one classifier produces the same highest score, such as galaxy and star both scoring 0.7, tie-breaking strategies have to be applied. These can include applying a fixed priority order based on prior knowledge or data distribution, randomly guessing, or using secondary classifiers.

The OvR approach has several benefits. Each individual classifier focuses on distinguishing one specific class from all others. This simplifies the learning task compared to directly solving a multi-class decision boundary. In theory, each binary problem can be trained on the full dataset, ensuring comprehensive exposure to all classes. For quantum implementations such as the QSVM or HHL-based SVM (HHL-SVM), however, training and executing multiple quantum circuits can still be resource-intensive. The number of classifiers scales linearly with the number of classes, increasing the total quantum kernel evaluations or quantum linear system solutions required. Despite this, the OvR approach remains a robust and widely-used strategy for multi-class classification. It is especially appealing when classes have distinctive features that can be readily separated from the rest, and scales more favourably as compared to the OvO approach. OvO requires a SVM for each pair of classes.

\subsection{Two-step SVM}

To mitigate the cost of requiring multiple SVMs for multi-class classification even more, we explored a two-step classification framework that uses only two SVMs to perform hierarchical decision-making. The two-step approach operates by first selecting one of the three classes to isolate in an initial binary classification. For example, the first SVM may be trained to distinguish between galaxies and non-galaxies (i.e., stars and QSOs grouped together). Once this initial decision is made, the second SVM is used to distinguish between the two remaining classes within the ``non-galaxy'' group. In this case, stars versus QSOs. In total, only two SVMs are trained and evaluated, making the method computationally more efficient than the full OvR or OvO approaches.

This hierarchical method is particularly well-suited for QML implementations such as the QSVM or the HHL-based SVM (HHL-SVM), where reducing the number of quantum kernel evaluations or quantum linear system solutions significantly decreases resource demands. Since quantum circuits are constrained in depth, width, and repetition (due to noise and decoherence), minimizing the number of classification passes is not only a cost-saving measure but also a hardware-aligned strategy.

A limitation of the two-step structure is the possibility of error propagation. If the first SVM misclassifies an input (e.g., predicting a galaxy as non-galaxy), the sample is routed incorrectly to the second SVM, which is not trained to handle galaxy samples. This leads to compounded errors that cannot be corrected downstream. If the first classifier is designed to handle the most separable class (for instance, QSOs may be easier to distinguish), then the likelihood of early misclassification can be reduced.

The two-step approach provides a practical trade-off between performance and quantum resource efficiency. It allows for targeted optimization of each binary decision boundary and limits the number of expensive quantum operations needed for training and inference, while still supporting structured multi-class classification in higher-dimensional, noisy datasets such as those encountered in astrophysics.

\subsection{Scaling}

To compare the computational scaling of the one-vs-rest (OvR) and two-step hierarchical SVM strategies, we estimate the number of kernel executions required for training and prediction for QSVM as a function of the dataset size. We plot the number of required gate operations for the HHL LS-SVM, counted during runs. The number of gate operations required for QSVM can be extrapolated from the kernel executions by multiplying by twice the number of features.

For QSVM, we fixed the number of test samples $M$ to be equal to the number of training samples $N$ for simplicity. Both are set to 300 samples each, as this is the number of samples used during training and testing. The training and prediction costs are given by:

\begin{equation}
T_{\mathrm{OvR}}(N) = C\times N^2, \quad 
P_{\mathrm{OvR}}(M,N) = C\times M\times N,
\end{equation}

\noindent with the total executions given by:

\begin{equation}
E_{\mathrm{OvR}}(N) = T_{\mathrm{OvR}}(N) + P_{\mathrm{OvR}}(N,N).
\end{equation}

For the two-step approach, the first classifier is trained on all $N$ samples:

\begin{equation}
T_1(N) = N^2.
\end{equation}

The second classifier is trained only on the subset of samples that do not belong to the class isolated in Step 1, or to the samples that were not classified as belonging to that class. Let $p_c$ be the proportion of class $c$ in the dataset; then the second step receives:

\begin{equation}
N_2 = (1 - p_c)N
\end{equation}

\noindent samples, with training cost:

\begin{equation}
T_2(N_2) = N_2^2.
\end{equation}

The total training cost is:

\begin{equation}
T_{\mathrm{TwoStep}}(N,p_c) = N^2 + \big[(1-p_c)N\big]^2.
\end{equation}

Assuming perfect first-stage classification, meaning no mistakes were made in the first SVM, the prediction cost equals the training cost, so the total executions are:

\begin{equation}
E_{\mathrm{TwoStep}}(N,p_c) = 2\,T_{\mathrm{TwoStep}}(N,p_c).
\end{equation}

We evaluate $E_{\mathrm{OvR}}$ and $E_{\mathrm{TwoStep}}$ over a range of samples, using the class proportions:

\begin{equation}
p_{\mathrm{GALAXY}} = 0.71, \quad 
p_{\mathrm{STAR}} = 0.17, \quad 
p_{\mathrm{QSO}} = 0.12.
\end{equation}

For each $N$, the kernel execution counts are computed according to the above formulas and plotted in \ref{fig:fig3}. The curves for each class in the two-step method differ according to the corresponding $p_c$, illustrating how class imbalance impacted computational cost under the perfect first-stage assumption. In reality the curves would deviate slightly.

The main advantage of using the HHL appears when analyzing the scaling of its circuits. In addition to requiring only a single execution for the training process, the indirect representation of the data via the kernel matrix allowed for the processing of larger datasets without increasing the execution time.

Because all circuits would only need to be executed once, the number of gate operations was analyzed instead. The number of operations for each circuit was calculated by creating each circuit with \texttt{Qiskit} and returning its size, with the QFT being the only part decomposed since it would appear as a single gate otherwise. 

\section{Results}\label{Results}

\begin{table*}[t]
\centering
\caption{Two-step classification performance results. Each class, indicated in the class column, is isolated separately in its own two-step hierarchical runs. The first SVM is a one-vs-rest and the second SVM is a standard one-vs-one to distinguish between the remaining classes. Values are reported as mean~$\pm$~standard error using macro (M) averages. HHL LS-SVM results have no uncertainty, as the results are deterministic.}
\label{tab:results_two}
\begin{tabular}{llcccc}
\toprule
\toprule
\textbf{Method} & \textbf{Class} & \textbf{Accuracy} & \textbf{Precision (M)} & \textbf{Recall (M)} & \textbf{F1-score (M)} \\
\midrule
\midrule
\multirow{3}{*}{QSVM} 
  & Galaxy & $0.969 \pm 0.004$ & $0.967 \pm 0.003$ & $0.930 \pm 0.009$ & $0.947 \pm 0.006$ \\
  & QSO    & $0.968 \pm 0.002$ & $0.973 \pm 0.003$ & $0.933 \pm 0.006$ & $0.950 \pm 0.000$ \\
  & Star   & $0.970 \pm 0.003$ & $0.977 \pm 0.003$ & $0.933 \pm 0.009$ & $0.953 \pm 0.009$ \\
\midrule
\multirow{3}{*}{CSVM} 
  & Galaxy & $0.968 \pm 0.001$ & $0.967 \pm 0.009$ & $0.940 \pm 0.006$ & $0.950 \pm 0.000$ \\
  & QSO    & $0.962 \pm 0.001$ & $0.957 \pm 0.003$ & $0.930 \pm 0.000$ & $0.940 \pm 0.000$ \\
  & Star   & $0.951 \pm 0.001$ & $0.927 \pm 0.007$ & $0.920 \pm 0.000$ & $0.923 \pm 0.003$ \\
\midrule
\multirow{3}{*}{HHL LS-QSVM} 
  & Galaxy & $0.893$& $0.908$ & $0.752$ & $0.812$  \\
  & QSO    & $0.754$ & $0.757$ & $0.653$ & $0.547$  \\
  & Star   & $0.81$ & $0.65$ & $0.67.12$ & $0.582$  \\
\midrule
\multirow{3}{*}{HHL LS-CSVM} 
  & Galaxy & $0.914$& $0.886$ & $0.8605$ & $0.872$  \\
  & QSO    & $0.841$ & $0.755$ & $0.727$ & $0.677$  \\
  & Star   & $0.894$ & $0.820$ & $0.8168$ & $0.818$  \\
\bottomrule
\bottomrule
\end{tabular}
\end{table*}

\begin{table*}[t]
\centering
\caption{The global one-vs-rest results and the three two-class binary SVM results. In the three binary SVMs, each class is isolated separately and classified against the rest. Average performance across three runs is given. Values are reported as mean~$\pm$~standard error using macro (M) averages. HHL LS-SVM results have no uncertainty, as the results are deterministic.}
\label{tab:results_ovr}
\begin{tabular}{llcccc}
\toprule
\toprule
\textbf{Method} & \textbf{Case} & \textbf{Accuracy} & \textbf{Precision (M)} & \textbf{Recall (M)} & \textbf{F1-score (M)} \\
\midrule
\midrule
\multirow{4}{*}{QSVM}
  & One-vs-Rest & $0.963 \pm 0.007$ & $0.963 \pm 0.007$ & $0.963 \pm 0.007$ & $0.963 \pm 0.007$ \\
  & QSO-vs-All  & $0.971 \pm 0.001$ & $0.937 \pm 0.007$ & $0.923 \pm 0.006$ & $0.933 \pm 0.006$ \\
  & Star-vs-All & $0.969 \pm 0.002$ & $0.980 \pm 0.000$ & $0.907 \pm 0.008$ & $0.937 \pm 0.007$ \\
  & Galaxy-vs-All & $0.958 \pm 0.001$ & $0.957 \pm 0.007$ & $0.937 \pm 0.006$ & $0.950 \pm 0.000$ \\
\midrule
\multirow{4}{*}{CSVM}
  & One-vs-Rest & $0.957 \pm 0.013$ & $0.960 \pm 0.010$ & $0.917 \pm 0.027$ & $0.933 \pm 0.019$ \\
  & QSO-vs-All  & $0.977 \pm 0.002$ & $0.957 \pm 0.003$ & $0.933 \pm 0.003$ & $0.943 \pm 0.003$ \\
  & Star-vs-All & $0.979 \pm 0.005$ & $0.987 \pm 0.003$ & $0.937 \pm 0.012$ & $0.960 \pm 0.009$ \\
  & Galaxy-vs-All & $0.971 \pm 0.009$ & $0.973 \pm 0.009$ & $0.957 \pm 0.014$ & $0.963 \pm 0.010$ \\
\midrule
\multirow{4}{*}{HHL LS-QSVM}
  & One-vs-Rest  & $0.808$ & $0.557$ & $0.627$ & $0.518$ \\
  & QSO-vs-All & $0.793$ & $0.680$ & $0.874$ & $0.699$ \\
  & Star-vs-All & $0.851$ & $0,704$ & $0.623$ & $0.515$ \\
  & Galaxy-vs-All & $0.900$ & $0.962$ & $0.829$ & $0.862$ \\
\midrule
\multirow{4}{*}{HHL LS-CSVM}
  & One-vs-Rest & $0.922$ & $0.891$ & $0.883$ & $0.884$ \\
  & QSO-vs-All & $0.883$ & $0.750$ & $0.912$ & $0.796$ \\
  & Star-vs-All & $0.894$ & $0.820$ & $0.817$ & $0.818$ \\
  & Galaxy-vs-All & $0.917$ & $0.904$ & $0.887$ & $0.895$ \\
\bottomrule
\bottomrule
\end{tabular}
\end{table*}

The two-step and one-vs-rest results in Tables \ref{tab:results_two} and \ref{tab:results_ovr} tell a similar story. In general, the well-known QSVM approach, performs better in every case. HHL LS-SVM is competitive in certain cases, but has clear drawbacks. This result is likely a combined effect of using explicit data point representation and clearly using class weights in the fitting process. The point was to compare the HHL LS-SVM implemented with the full dataset, but with a limited number of representative elements, to the QSVM with much fewer samples, but with explicit data point representation. 

There are some outlier observations to take note of. The two-step hierarchical results in Table \ref{tab:results_two} indicate that the most competitive results are observed for the HHL LS-SVM for the case where galaxies are isolated in the first binary SVM. The second best results are seen when isolating blue stars. This means that the HHL LS-SVM struggles with QSOs the most. This is expected, since QSOs make up the smallest minority in the dataset. Galaxies are the majority class, which means that their signatures are learned more easily. Stars are less distinct from galaxies and form the second minority class, which explains why the model performs worse in comparison with galaxies, but performs better in comparison with QSOs. It is harder for any model to learn the signature of a specific class the less samples there are available to train it on. This suggests that it is important to consider domain knowledge, if any, when deciding which class to isolate first in the two-step approach fo. 

Comparison of QSVM and CSVM results directly, shows similar observations to what is known \cite{slabbert2024pulsar, slabbert2024hybrid}: The results are similar, with the classical CSVM edging out slightly overall. Both methods were fitted using class weights and all ten selected features. QSVM achieved slightly superior performance for the case where stars are isolated, as well as for the main one-vs-rest result. This is a direct contradiction from what was observed for LS-SVM. The most obvious explanation for this is that the class weights in the fitting process mitigate the minority/majority problems during training.

Comparison of the classical and quantum HHL LS-SVM is not that different from the QSVM and CVSM comparison. The results are close in most cases, with the classical method edging out in general, which is explained by a small variation in values that occurs when implementing the kernel matrix using the quantum phase estimation.

\begin{figure*}[t]
    \centering
    \includegraphics[width=\textwidth]{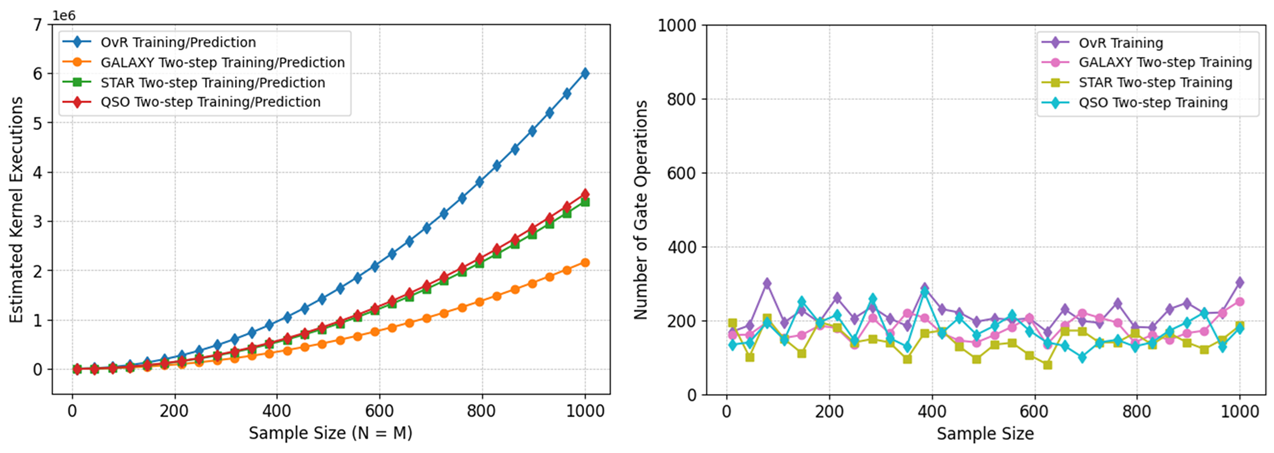}
    \caption{Scaling plots for QSVM (left) and the quantum HHL LS-SVM (right). To convert estimated kernel executions to number of gate operations, simply multiply by the number of features times two (10 features times 2). Kernel executions for QSVM are given instead of gate operations, since the scaling is already obviously worse than the HHL method and to keep the y-range down as much as possible. Scaling for QSVM is quadratic because it requires and evaluation for each pair-wise combination of feature vectors. $ N = M$ indicates that the number of training and test samples are equal. Classes with less samples making up the isolated class in the first SVM in the two-step approach, will lead to more samples in the second SVM, ultimately requiring more executions overall. The HHL LS-SVM shows a clear constant trend in the required number of gate operations, with some fluctuation.}
    \label{fig:fig3}
\end{figure*}

It should be clear from the scaling plot in Figure \ref{fig:fig3} that the HHL LS-SVM scales much more favourably compared to the QSVM. The QSVM scales quadratically with the number of samples, and because both the two-step and one-vs-rest approaches require more than one SVM, this causes an even worse scaling overall. The HHL LS-SVM scales much more favourably with a clear, essentially constant trend. This happens because the data was processed classically and did not interfere directly with the circuit size, maintaining the quantum resources constant between different samples sizes. This makes the HHL LS-QSVM the obvious choice if the number of samples is large. It is important to note that, since the QSVM uses explicit data point representation in the form of encoded features instead of the indirect use in the HHL, the QSVM has an advantage when dealing with more complex datasets.
It is important to note that QSVM also scales with the number of features, because more features require more qubits to encode all the information.

Real runs for the HHL LS-SVM were attempted, but the amount of noise introduced significantly reduced the ability of the method to classify samples accurately. It is known that the HHL algorithm is really sensitive to noise, especially at higher dimensions, which makes it a challenge to execute with the high rate of noise observed in current quantum devices. QSVM was not run on real devices, as each kernel value calculation is a query to the device. Although noise is as limited as it can be with just two gates in sequence, the query limitation was difficult to work with during the NISQ era, since each circuit execution runs at the order of seconds.

\section{Conclusion}\label{Conclusion}

We applied both the QSVM and HHL LS-SVM to a feature-selected version of the \texttt{SDSS} dataset. Training the QSVM involved standard \texttt{scikit-learn} SVM fitting that included class weights. The class weights were calculated using the class distribution. The choice of quantum kernel used y-rotations applied to ten qubits, where the rotational parameter was equal to one of the ten features, scaled to be between $[0, \pi]$. The second unitary operator was the complex conjugate transpose of the first, only using a different feature vector, and the kernel value was calculated by a projective measurement using the zero-state density matrix and z-expectation value measurement in \texttt{Pennylane}. Training of the HHL involved classically processing the data to generate a linear system to be solved by the HHL algorithm. The solution was a LS-SVM. Performance was assessed using standard accuracy and macro averages of recall, precision, and F1-score. Two line plots showcasing the scaling of the number of gate operations required to run each method are also given for clear comparison in the context of computational costs. No real device runs are included, as noise significantly worsens the results for the HHL algorithm, and the runtimes for QSVM would be too high.

The results show a consistent pattern across the chosen methods: QSVM outperforms HHL LS-SVM, likely due to the use of direct feature representation and the integration of class weights during training, while HHL LS-SVM is limited to a small number of representative elements. The two-step hierarchical approach highlights that isolating the majority class galaxy first produces the most competitive results, followed by stars, whereas QSOs are consistently misclassified because of their minority status. Similar trends are observed in the one-vs-rest approach, with the galaxies classification remaining the most reliable. Classical SVMs (CSVM) perform slightly better than QSVM overall, as expected, but there are some exceptions for specific cases. Comparisons between classical and quantum HHL LS-SVM show small differences, with failures occurring mainly when QSOs are involved. Scaling analysis reveals a clear trade-off: QSVM scales quadratically with the number of samples and also scales with the number of features, while HHL LS-SVM scales essentially constantly, making it favourable for large datasets at the cost of limited representation capacity and sensitivity to noise. This noise prevented successful real-device runs. Overall, the results suggest that careful consideration of class characteristics, hierarchical strategy is crucial when applying HHL LS-SVM, whereas QSVM and CSVM offer a more robust and straightforward approach.

Once we move beyond the NISQ era, running these methods using real devices will be the obvious next step. This will give a true perspective on what the performance and runtime expectations will be on hardware. Another way to improve the QSVM, would be to remove the class weights from the comparison or to parameterize the quantum kernel for training \cite{slabbert2025spectral, xu2024quantum, rodriguez2025neural, hubregtsen2022training}. The HHL LS-SVM can be improved by using more complex kernel matrices and using more powerful simulators to expand the number of representative elements used. The point was to bias the comparison in favour of the well-established QSVM, since the highest possible performance is typically the goal. These comparisons show that while QSVM remains the better all-round performer, the fact that HHL LS-SVM can remain competitive under such constraints is a promising sign that motivates its use.

\bmhead{Acknowledgments}
We acknowledge Gabriel Pinto and Anna Scaife from the University of Manchester for their suggestion of the \texttt{SDSS} dataset.

\subsection*{Funding}

This work was financed in part by the Coordenação de Aperfeiçoamento de Pessoal de Nível Superior - Brasil (CAPES) - Finance Code 001.
This work was funded by the South African Quantum Technology Initiative (SA QuTI) through the Department of Science, Technology, and Innovation of South Africa.

\subsection*{Conflict of interest}

The authors declare no conflict of interest.

\section*{Data Availability Statement}

The reduced \href{https://zenodo.org/records/3768398}{\texttt{SDSS}} dataset is available for download \cite{clarke2020identifying, a_o_clarke_2020_3768398}.

\subsection*{Authors' contributions}

G.P.C. and D.S. performed the numerical experiments and analyses, compiled the results, and wrote the manuscript. D.S. was responsible for the implementation of the QSVM and G.P.C. for the HHL LS-SVM. F.P. and L.K. supervised the research. All authors reviewed and discussed the analyses and results and contributed to editing the manuscript.

\bibliography{references}

\end{document}